\documentclass{v16nufact}

\def\mco{\multicolumn}

\def\be{\begin{equation}}
\def\ee{\end{equation}}
\def\bea{\begin{eqnarray}}
\def\eea{\end{eqnarray}}

\newcommand{\Photo}{}

\begin{document}
\vspace*{4cm}
\title{Neutrino Monte-Carlo Event Generators and Cross-section Data}

\date{\today}
\author{P.~Stowell$^{1}$, S.~Cartwright$^{1}$, L.~Pickering$^{2}$, C.~Wret$^{2}$, C.~Wilkinson$^{3}$ \\}
	
\address{$^{1}$University of Sheffield, $^{2}$Imperial College London, $^{3}$University of Bern}

%\affiliation[a]{University of Sheffield, Department of Physics and Astronomy, Sheffield, United Kingdom}
%\affiliation[b]{Imperial College London, Department of Physics, London, United Kingdom}
%\affiliation[c]{University of Bern, Albert Einstein Center for Fundamental Physics, Laboratory for High Energy Physics (LHEP), Bern, Switzerland}

\maketitle

\abstracts{ 
In recent years a number of new theoretical models have been implemented into Monte-Carlo neutrino interaction event generators. Being able to compare multiple model predictions is invaluable to the field, especially as it is unfortunately still unclear which one provides the best fit to the entire collection of neutrino cross-section data. Using a recently developed neutrino generator tuning framework (NUISANCE), we review a selection of models in the NEUT and NuWro generators through comparisons to existing bubble chamber, MINERvA, and MiniBooNE cross-section data. 
}

\section{Introduction}
There are currently multiple neutrino event generators available, providing a large range of different interaction models to choose from when trying to construct a complete nuclear scattering model. The NUISANCE framework \cite{nuisance} has recently been released to try and provide the neutrino community with the necessary tools to help select and tune these different generators by making comparisons to existing cross-section data. The structure of NUISANCE allows multiple generators to be read into the same analysis routines, enforcing consistency of the signal selections applied and ensuring any differences observed are due to the underlying physics assumptions in each model being compared. Providing an interface between generator reweighting engines and ROOT's minimiser libraries \cite{ref:ROOT} has created a powerful tool that can be used to automatically tune generator models to multiple datasets by scanning the parameter space and minimising a likelihood test statistic. In these proceedings preliminary tuning results of several different components of the NEUT \cite{neutgen} and NuWro \cite{nuwrogen} generator models are compared.

\section{Bubble Chamber Tunings}
Fermi motion and binding energy effects are small for deuterium targets. This allows good constraints to be placed on neutrino-nucleon interaction models through fits to deuterium-filled bubble chamber data. Cross-section and event rate distributions from charged-current (CC) neutrino quasi-elastic (QE) and charged pion production ($1\pi^{+}$) measurements from the ANL, BNL, BEBC, and FNAL experiments, were chosen for these studies \cite{ref:anlccqe,ref:anlccres1,ref:anlccres2,ref:bnlccqe,ref:bnlccres1,ref:bnlccres2,ref:fnalccqe,ref:bebcccqe}. The measured cross-section and event rate distributions were digitised and added as dataset comparison routines into the NUISANCE framework so the data could be included in joint likelihood fits of the generator models.

The nominal NEUT and NuWro free nucleon models for QE and $1\pi^{+}$ scattering of free nucleons were chosen as candidate models to be tuned with the NUISANCE framework. These generators both use the Llewellyn-Smith\cite{ref:llewellynsmith} (LS) model to describe quasi-elastic scattering, and the Rein-Sehgal\cite{reinseghal} (RS) model to describe pion production, their main difference being that NEUT simulates multiple nuclear resonances using the RS model, whereas NuWro simulates only the $\Delta(1232)$ component, relying on a $\Delta$/DIS extrapolation to populate the higher order resonances.
Both generators have ``reweight engines'' allowing the user to make model predictions over a range of model parameters after event generation \cite{nubros}. These reweight engines were used to study variations in a set of free model parameters. For the QE model only the quasi-elastic axial mass parameter ($M_{A}^{QE}$) was treated as free. In the pion production model both the resonant axial mass ($M_{A}^{1\pi}$), and the axial coupling constant ($C_A^5$) were treated as free.

The overall similarity between the two generators provides an additional validation test of the tuning results. Since both generators use the LS and RS models it is expected they should obtain similar best fit results when tuning to distributions where the majority of events originate from low hadronic mass events ($W < 1.4$ GeV). The published flux distributions were used to generate charged-current events in each generator. The target in each case was considered to be a free proton and neutron to give a combined deuteron cross-section without binding energy or Fermi motion effects. From these Monte-Carlo (MC) samples, events were selected that matched the published signal selections and normalised to give matching cross-section predictions. In the cases where only event rate information was given the predictions were normalised to match the integrated event rate in the data. An additional correction was applied to the QE model predictions to convert them from free nucleon predictions to that for a bound deuteron \cite{ref:singh}. This correction, applied as a function of true $Q^{2}$, was found to have a negligible effect on the fits but was left in to maintain consistency with tuning studies shown in the original publications \cite{ref:anlccqe}.

\begin{table}[t]
\caption[]{
\label{tab:tuningbc}
Tuning results for the free nucleon interaction models in the NEUT and NuWro Monte-Carlo generators.
}
\begin{center}
\begin{tabular}{| c | c c | c c c |}
\hline
   & \mco{2}{|c|}{Quasi-elastic} & \mco{3}{|c|}{Resonance} \\
\hline
Model & $M_{A}^{QE}$ (GeV) & $\chi^{2}$/DOF & $M_{A}^{1\pi}$ (GeV) & $C_{A}^{5}$ & $\chi^{2}$/DOF \\
\hline
NEUT 5.3.6 & $1.04 \pm 0.03$ & 159.8 / 146 & $0.89 \pm  0.04$ & $1.02 \pm 0.05$ & 102.8 / 102 \\
NuWro v12 & $1.03 \pm 0.03$ & 154.4 / 146 & $0.92 \pm  0.04$ & $1.04 \pm 0.05$ & 111.2 / 102 \\
\hline
\end{tabular}
\end{center}
\end{table}

\begin{figure}
\begin{center}
\includegraphics[width=0.4\textwidth]{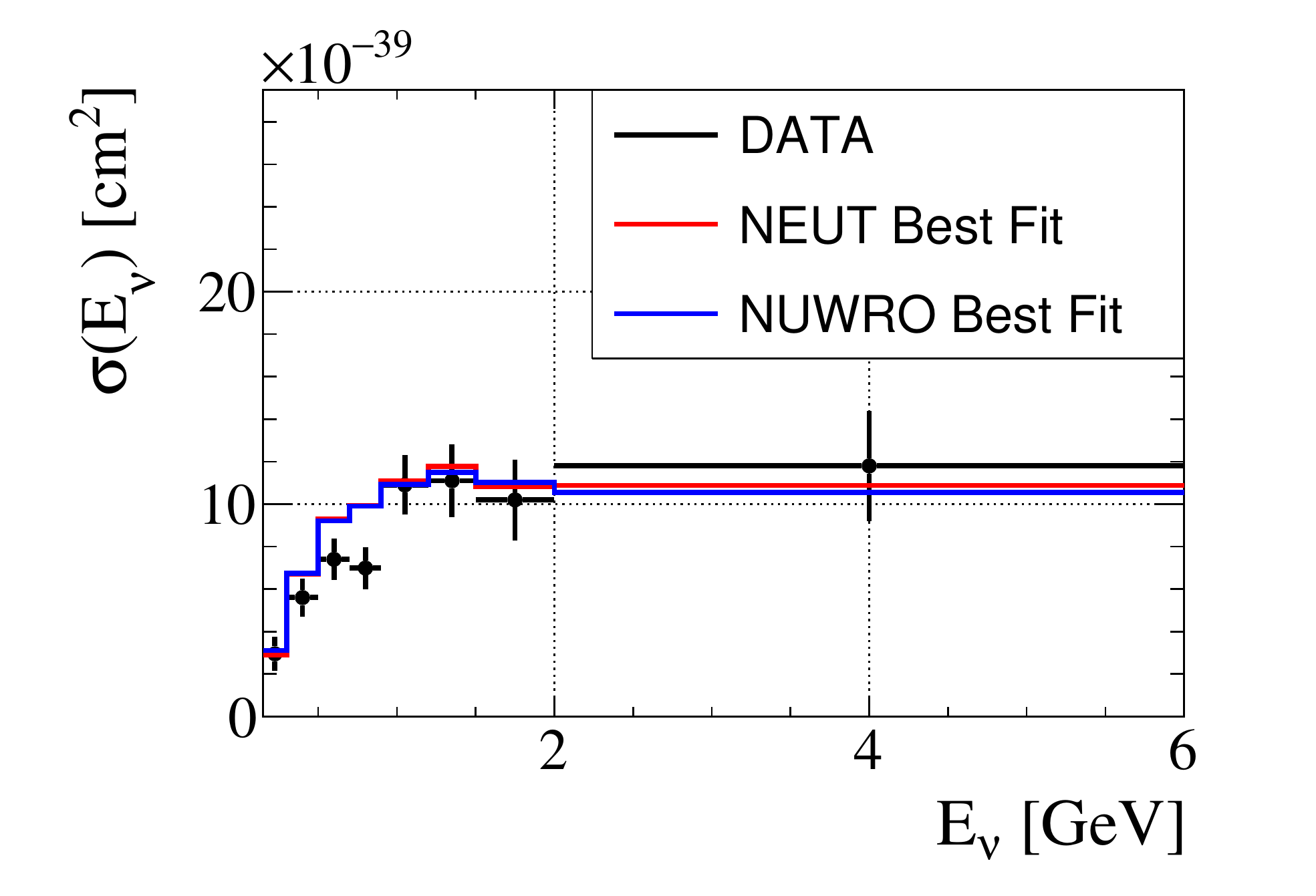}
\includegraphics[width=0.4\textwidth]{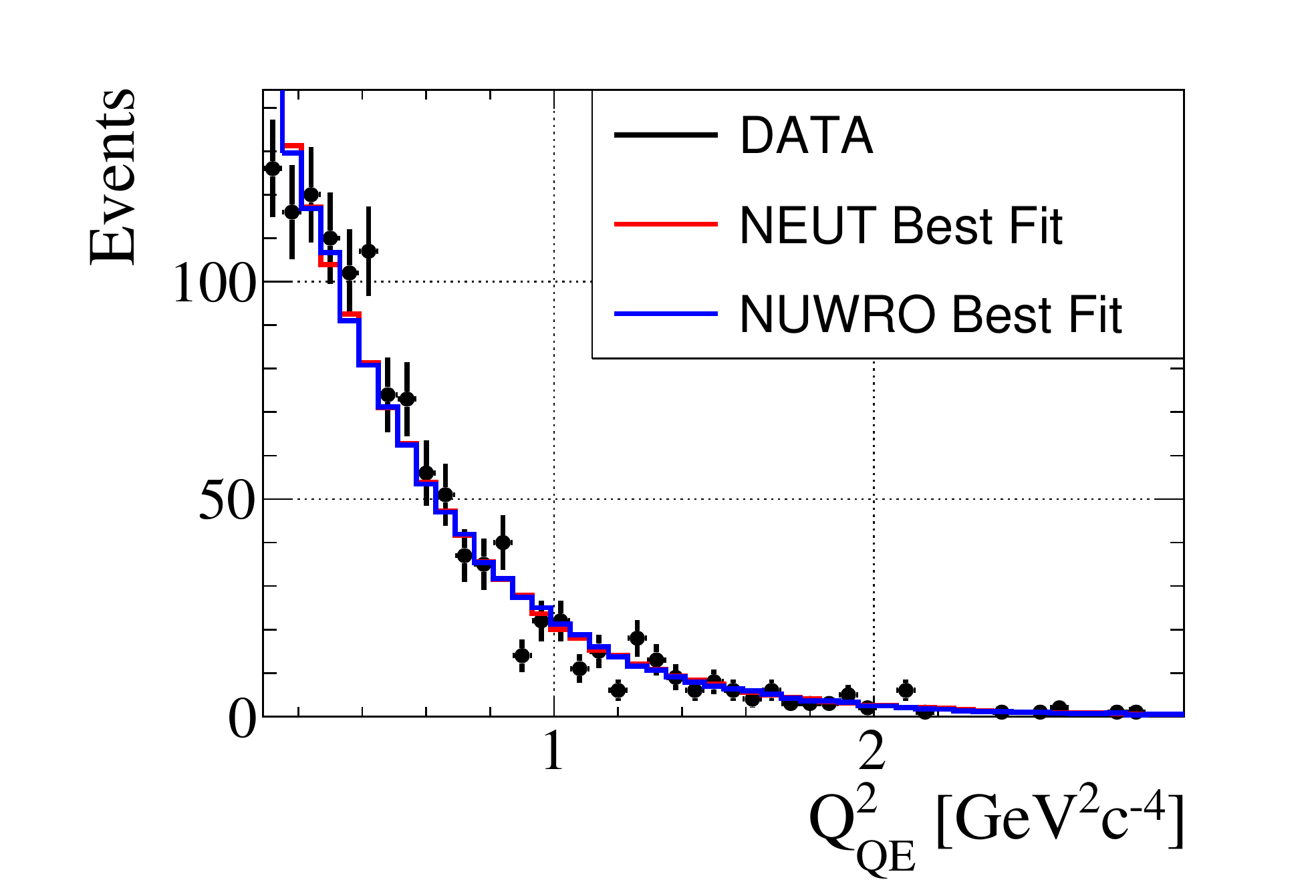}
\caption[]{\label{fig:ccqebc}Comparisons of the best fit predictions in the NEUT and NuWro generators to CCQE data. (left) ANL $E_{\nu}$ cross-section data used to constrain the ANL CCQE normalisation. (right) BNL $Q^{2}$ event rate data used to place an additional shape constraint on the BNL cross-section predictions.
}
\end{center}
\end{figure}

\begin{figure}
\begin{center}
\includegraphics[width=0.4\textwidth,trim=0 0 0 30, clip]{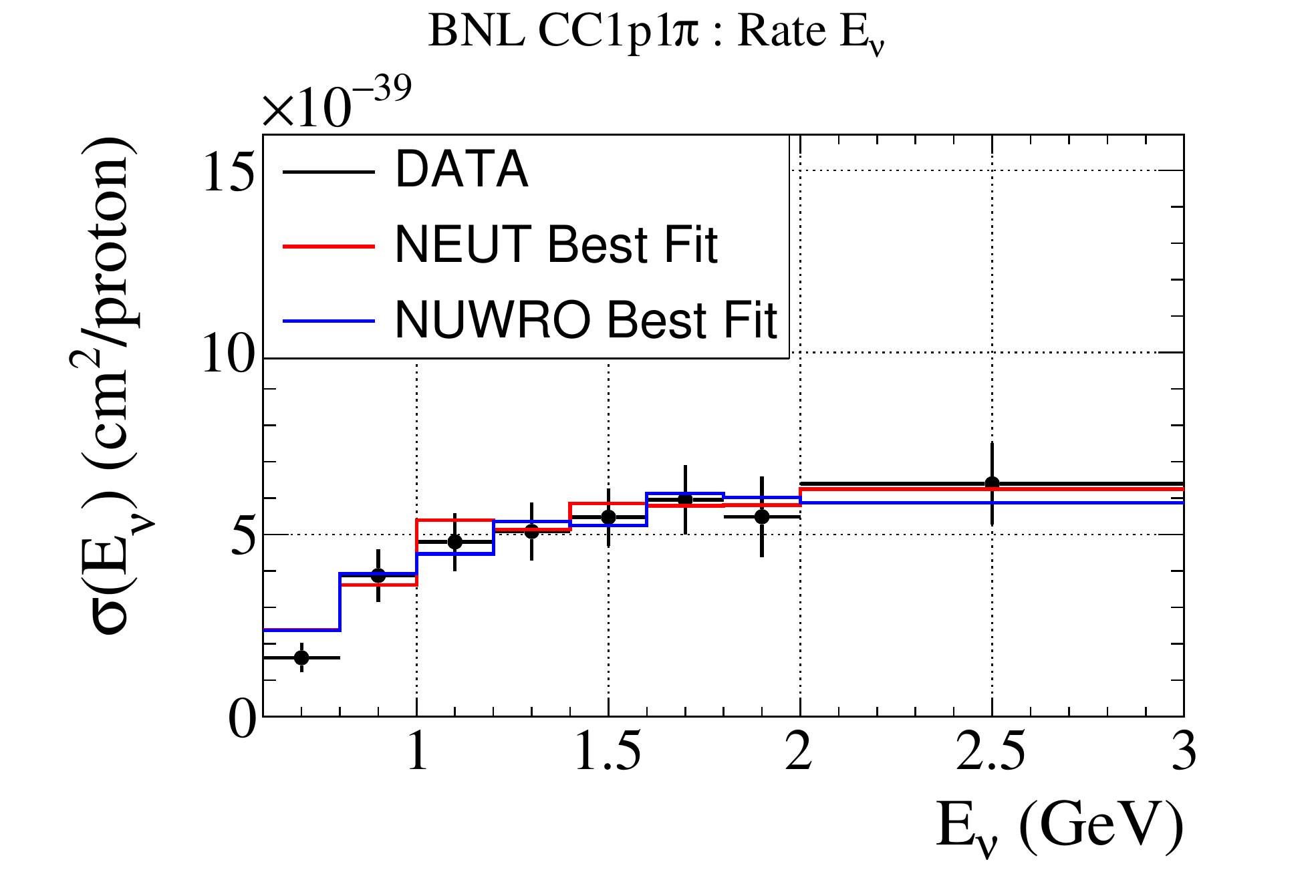}
\includegraphics[width=0.4\textwidth,trim=0 0 0 30, clip]{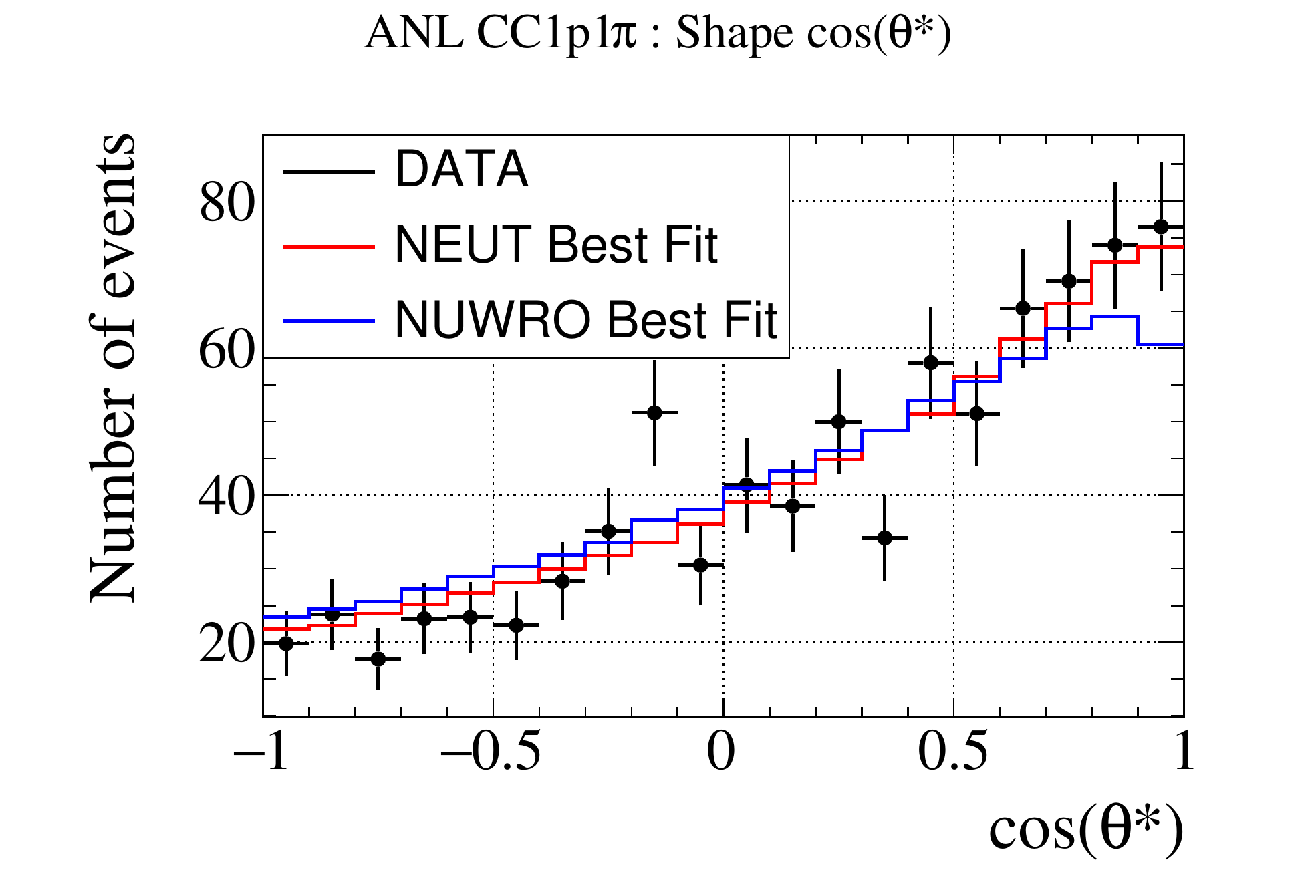}
\caption[]{\label{fig:ccresbc} Comparisons of the best fit predictions in the NEUT and NuWro generators to CC1$\pi$ data. (left) BNL $E_{\nu}$ cross-section data used to constrain the BNL CC1$\pi$ normalisation (right) ANL Adler Angle event rate data used to place an additional shape constraint on the BNL cross-section predictions. Effects of higher order resonances can be seen in the difference between NEUT and NuWro at low angles. }
\end{center}
\end{figure}

A joint likelihood was formed within the NUISANCE framework by first selecting a single distribution from each publication to place a constraint on the normalisation for that measurement (e.g. CCQE $\sigma(E_{\nu}))$. An additional shape-only likelihood was then added for each remaining distribution in the measurement itself (e.g. CCQE $Q^{2}$ Event Rates) to form a total likelihood for that measurement. The purpose of adding these shape-only terms is to minimise any bias that may be introduced when a model is tuned to only a single distribution, whilst trying to avoid issues with over-counting placing too strong a constraint on the total cross-section. These likelihoods for each dataset were added uncorrelated to form a total likelihood for the chosen models in the study.
The NUISANCE tuning framework was set up to automatically scan the parameter space until a best fit parameter set was found.  The results can be seen in Table \ref{tab:tuningbc}, with examples of the best fit predictions for both generators in Figs.\ \ref{fig:ccqebc} and \ref{fig:ccresbc}. Both generators were found to be capable of describing the data with an acceptable goodness-of-fit. The disagreement seen between the generators in the quasi-elastic $\chi^2$ results are due to slight differences in the generated MC statistics, whereas the differences in the $1\pi^{+}$ fits arise from fundamental differences in the generator models themselves. This can be seen in Fig. \ref{fig:ccresbc} where higher order resonances can be seen contributing to the NEUT prediction at high angles introducing a difference between the two generator predictions. In both cases the differences are not large enough to significantly shift the tuning results, with both generators finding best fit results in agreement with one another, providing a set of free nucleon parameters suitable for propagation to future nuclear tuning studies.

\section{NuWro/NEUT LFG Tunings}
When extending neutrino interaction models to nuclear targets, an inclusive generator model must also consider how the presence of the nuclear medium can modify the interaction. We consider the latest model available in the NEUT 5.3.6 generator, consisting of a Relativistic Fermi Gas\cite{smithmoniz} (RFG) with relativistic RPA correction and a Nieves multi-nucleon model \cite{nieves} (NEUT RFG+Nieves).  The choice of nuclear spectral function to model nucleon binding energy and Fermi motion introduces a problem in generator model tuning, since multiple discrete models are available. It is believed that any direct measurements of quasi-elastic scattering are also likely to be sensitive to additional multi-nucleon interaction channels (2p2h) that can produce final states of similar topologies to true quasi-elastic scattering inside the nucleus \cite{nieves}. 
For comparison we also consider two alternative models in the NuWro generator, a model with a local Fermi gas RPA correction and a Nieves 2p2h model (NuWro LFG+Nieves), and a RFG with a transverse enhancement model\cite{temmodel} (NuWro RFG+TEM). 

Each model was tested against MiniBooNE and MINERvA CCQE data in both neutrino and anti-neutrino runs \cite{ref:mbnu,ref:mbnub,ref:minnu,ref:minnub}. Although the collaborations define their signal as ``true CCQE interactions'', experience suggests that all four measurements are in fact sensitive to both CCQE and 2p2h interaction channels. Model predictions corresponding to each dataset were therefore produced by generating events with the published flux distribution and selecting only those events which originated from one of these two interaction channels. A joint sample likelihood for the study was defined using MiniBooNE $T_{\mu}-\cos\theta_{\mu}$ data with shape-only uncorrelated errors and a floating normalisation, and MINERvA $Q^{2}_{QE}$ data with full covariance between neutrino and antineutrino distributions, matching the method used in Ref.\ 2. To look at variations in both of the interaction channels, the quasi-elastic axial mass (alters only CCQE interactions) and 2p2h normalisation (alters only 2p2h interactions) were treated as free parameters that could could be changed to improve agreement between the data and MC.

The $\chi^{2}$ values shown in Table \ref{tab:ccqeresults} are unrealistically small, because the MiniBooNE 2D distribution public data release does not provide bin-to-bin correlations. 
When varying both parameters freely similar results were found for all three models, an inflation of the axial mass away from the bubble chamber tuning result, and a large suppression of the 2p2h cross-section normalisation compared to the nominal prediction. Both parameters were estimated to be highly correlated when using MINUIT's HESSE\cite{ref:hesse} routine to estimate parameter errors, a feature of the strong shape-constraint that the MINERvA dataset places on the fit. 
The use of a local Fermi gas model was insufficient to relieve the tensions seen in previous joint fit studies to this data and a significant model variation is likely needed to relieve the tensions whilst still maintaining consistency with other theoretical and experimental constraints. One significant problem with this method of tuning individual interaction channels to this data is that an unknown fraction of pion-less delta decay events was subtracted from the each distribution, directly by the MiniBooNE collaboration in their background subtraction procedure, and indirectly by MINERvA in their cut on their recoil energy deposited inside the detector outside the interaction vertex. If reliable constraints on free cross-section parameters for nuclear targets are to be extracted, a series of dedicated tuning studies using more inclusive signal definitions with minimal model-dependent background corrections is required.

\begin{table}[t]
\caption[]{\label{tab:ccqeresults} Tuning results for the NEUT and NuWro CCQE+2p2h models when compared in joint fits to MiniBooNE and MINERvA quasi-elastic cross-section data.}
\begin{center}
\begin{tabular}{| c | c c | c | c |}
\hline
Model & $M_A$ (GeV)& 2p2h Norm (\%) & $\chi^{2}$/DOF \\
\hline
NuWro LFG+Nieves & $1.16 \pm 0.03$ & $8.3 \pm 11.9$ & 100.74 / 229 \\
NuWro RFG+TEM  & $1.15 \pm 0.03$ & $21.3 \pm 12.5$ &   93.62 / 229 \\
NEUT RFG+Nieves & $1.14 \pm 0.03$ & $25.5 \pm 12.4$ & 106.25 / 229
\\
\hline
\end{tabular}
\end{center}
\end{table}

\begin{figure}
\begin{center}
\includegraphics[width=0.35\textwidth]{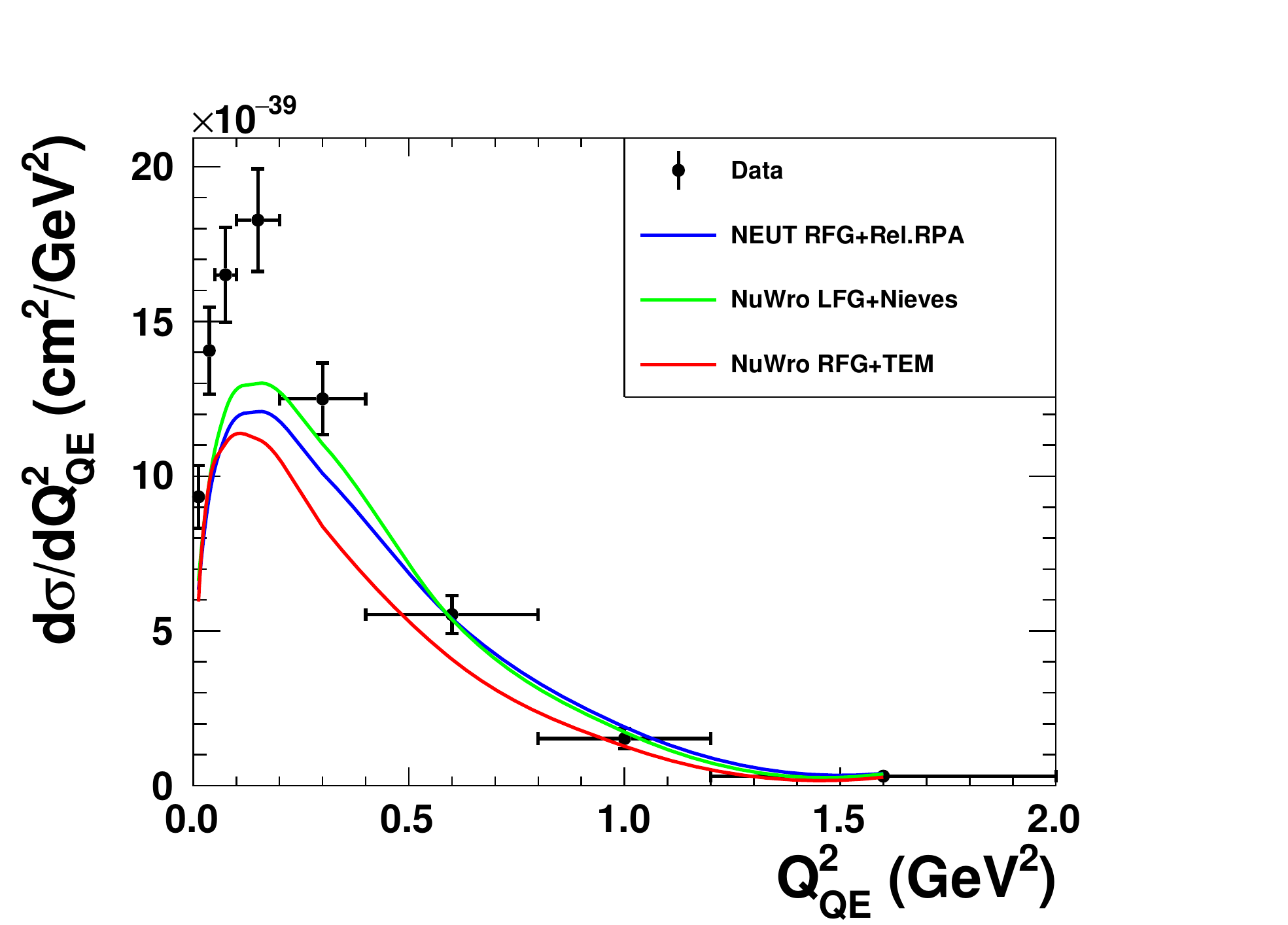}
\includegraphics[width=0.35\textwidth]{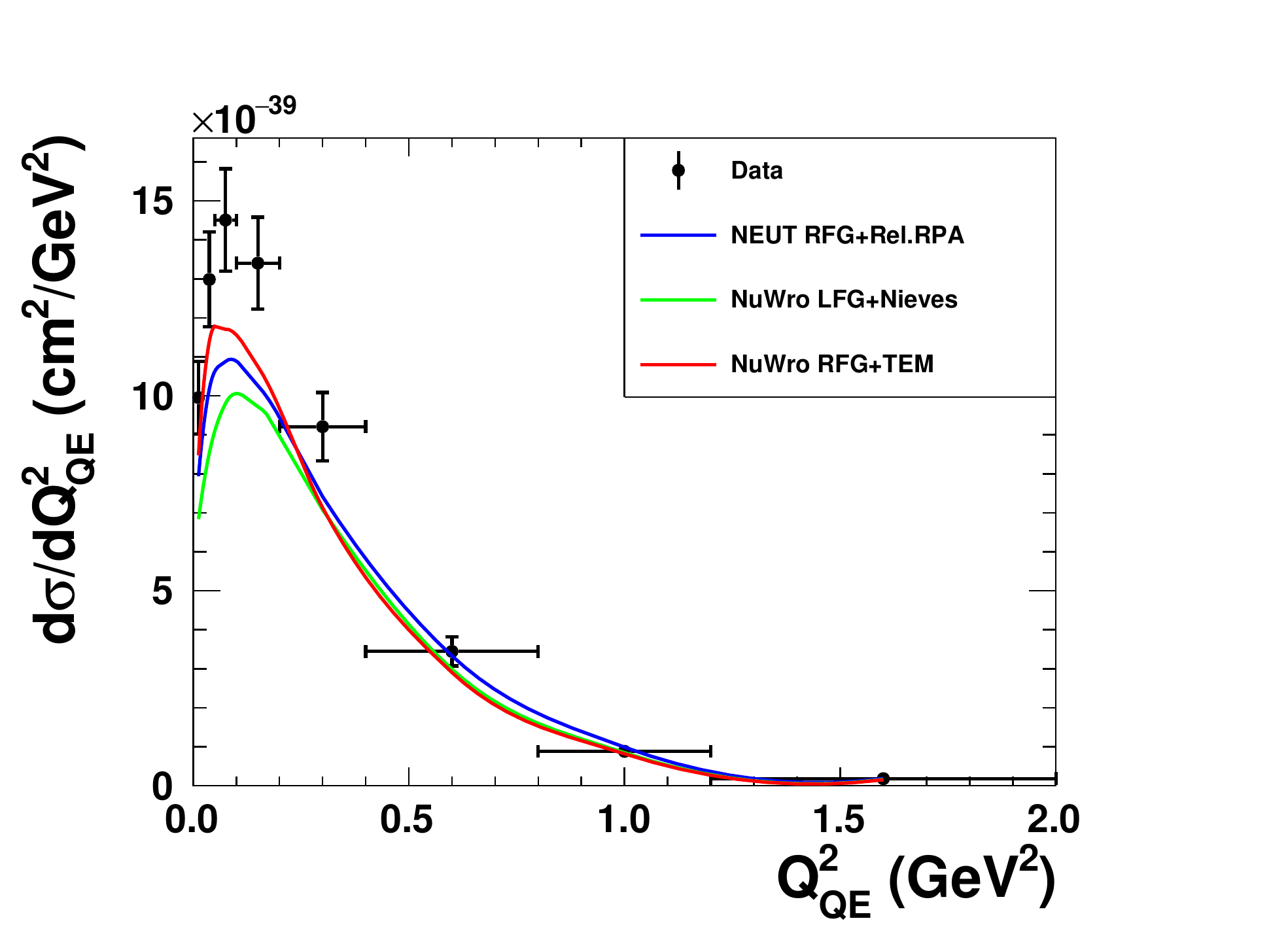}
\caption[]{Comparison of the best fit MC predictions in NEUT and NuWro to MINERvA CCQE data. The clear difference in the normalisation between the MC and data arises from the MINERvA data placing a much stronger constraint on the cross-section shape than its normalisation.}
\end{center}
\end{figure}

\section{MINERvA CC-inclusive comparisons}
The MINERvA collaboration has attempted to study the presence of nuclear effects in neutrino carbon interactions directly through the extraction of both the 3-momentum transfer ($q_3$) and hadronic recoil energy for a given event \cite{minervadata}. The variable ``energy available'' ($E_{av}$) is defined as the sum of kinetic energy of protons and charged pions, and the total energy of neutral pions, electrons, and photons, leaving the nucleus. Subtracting the muon energy from the observed energy deposited around the vertex allows a CC-inclusive event selection to be unfolded into a differential cross-section measurement in terms of $E_{av}$ and $q_{3}$. Comparisons between this data and GENIE have shown disagreement in the ``dip'' region at high $q_{3}$ between the quasi-elastic and resonance peaks ($0.4 < q_{3}/\mbox{GeV} < 0.6$) . Similar differences between model predictions and data have been observed by the NOvA collaboration when studying hadronic recoil energy, and it has been suggested that changes to how we model 2p2h interactions could relieve this tension.

\begin{figure}
\begin{center}
\includegraphics[width=0.68\textwidth]{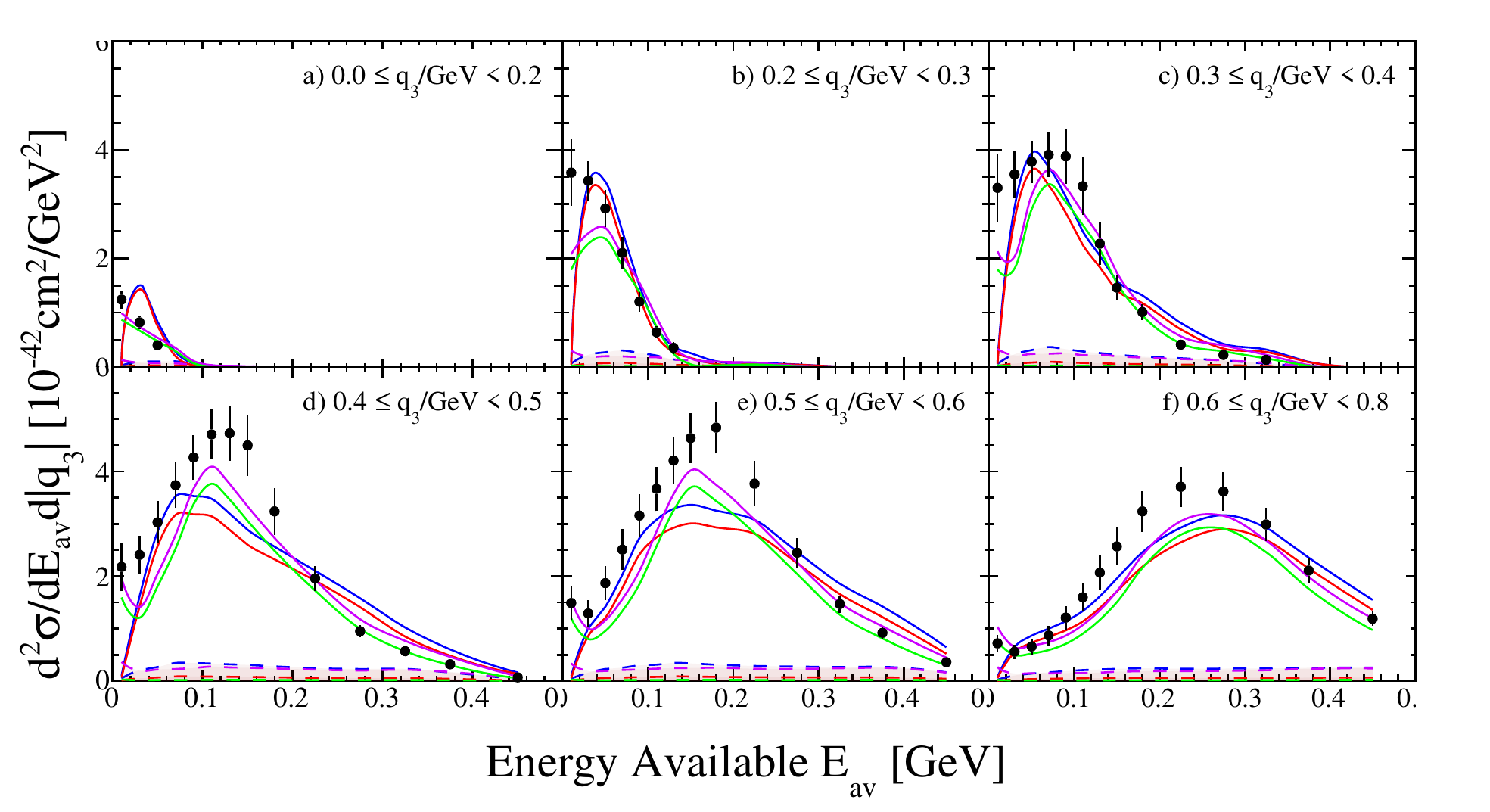}
\caption{\label{fig:minervaneutnuwro} Comparison of the previous NEUT and NuWro tuning results to MINERvA low recoil scattering data. Shown are the NEUT predictions using the bubble chamber tuning (blue), the NEUT predictions using the CCQE tuning (red), the NuWro  LFG+Nieves predictions with bubble chamber tuning (purple), and CCQE tuning (green). The dashed lines of matching colour show the predicted 2p2h contribution to the cross-section in each bin.}
\end{center}
\end{figure}

\begin{figure}
\begin{center}
\includegraphics[width=0.68\textwidth]{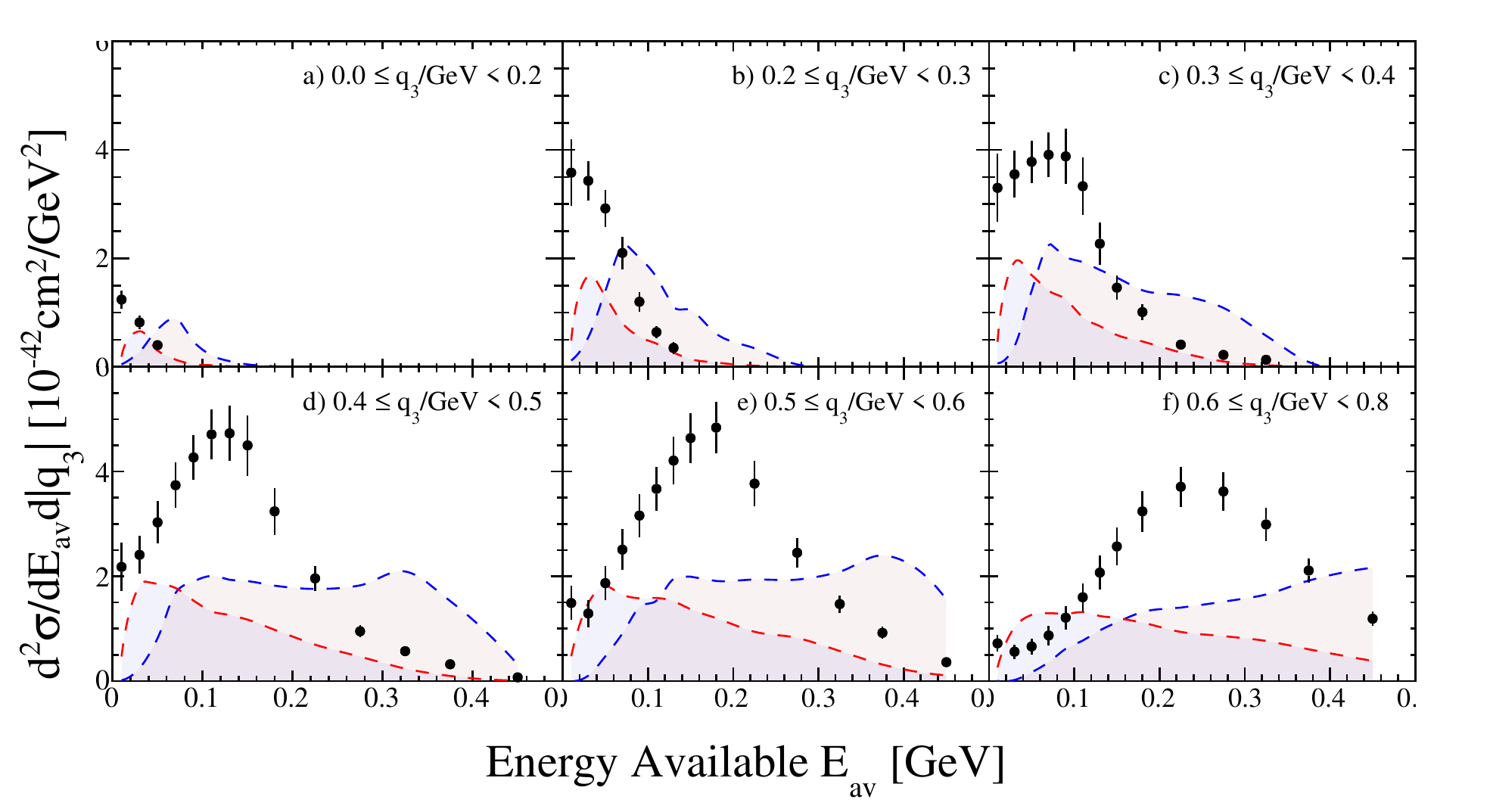}
\caption{\label{fig:neutpnnn} Comparison of NEUT 2p2h pp/nn pair contributions in the $E_{av}$ variable. Each prediction has been scaled up by a factor of 10 so that its shape is clear. Shown in red and blue are the nn and pn contributions respectively. }
\end{center}
\end{figure}

For comparison the best fit results from the NEUT and NuWro tunings to both bubble chamber and carbon measurements are compared to this ``recoil energy data'' in Fig.\ \ref{fig:minervaneutnuwro}. Simple variations in the axial mass and 2p2h normalisation are found to be incapable of filling in the disagreement between the data and MC, but the large shape disagreement in the ``dip'' region is significantly smaller for NuWro as a result of using a local Fermi gas model. Since the signal definition is  CC-inclusive and extremely sensitive to final state particle multiplicities, it is also possible to create similar predictions through multiple smaller variations of different features of the inclusive generator model. For example, Fig.\ \ref{fig:neutpnnn} shows the different contributions to the NEUT and  2p2h cross-section from pn and nn pairs. This fraction of these pairs currently has a reasonably large uncertainty assigned, and is just one of many examples of free parameters that could be used to sculpt the total CC-inclusive prediction to better match the data.

The major complication of trying to use such a measurement on its own to understand where models may be deficient is that final state ``recoil energy'' variables will be extremely sensitive to final state interaction models. Changes in these models can cause events to migrate in ``recoil energy'' space making it difficult to disentangle which exact features of the model may be problematic. 
 Since no measurements have been made in these kinematic variables in the past, it is difficult to tell in which of the many interaction channels or FSI model the tensions may really lie, and a full CC-inclusive model tuning with additional constraints from CC$0\pi$/CC$1\pi$ data may be required to extract reliable results from these recoil energy measurements.

\section{Conclusion}
The NUISANCE tuning group has defined a publicly accessible framework that supports neutrino event generator tuning. 
Early studies of the NEUT and NuWro generator free nucleon models have found that both generators obtain consistent results when tuning to deuterium bubble chamber data. One weakness of these tunings is the lack of correlations across  bubble chamber experiments, and future studies will look at correlating flux uncertainties in the studies to obtain more reliable best fit parameters.
When compared to nuclear CCQE data from MINERvA and MiniBooNE, both generators were also found to produce very similar tuning results despite much clearer differences between the models investigated. The use of alternative spectral function definitions was found to make little difference in the observed suppression of the 2p2h normalisation, hinting that the difficulties in achieving good agreement between these experiments may lie with the cross-section extraction methods used to obtain the data itself. Finally, comparisons of the fit results to MINERvA CC-inclusive showed that more significant modifications to the full cross-section model are required to obtain a reasonable agreement with newer CC-inclusive cross-section data.

\section*{Acknowledgments}
The author wishes to thank the UK STFC for supporting this work.


\section*{References}
\begin{thebibliography}{unsrt} 

\bibitem{nuisance}  P.~Stowell,~L.~Pickering,~C.~Wilkinson,~C.~Wret, 
``NUISANCE Framework'', http://nuisance.hepforge.org/

\bibitem{ref:ROOT} Brun, Rene, and Fons Rademakers. 
%"ROOT an object oriented data analysis framework." 
Nucl.\ Instrum.\ Meth.\ A {\bf 389.1} (1997)

\bibitem{neutgen} Y.~Hayato,
  %``A neutrino interaction simulation program library NEUT,''
  Acta Phys.\ Polon.\ B {\bf 40}, 2477 (2009).
  
\bibitem{nuwrogen}  C.~Juszczak,
  %``Running NuWro,''
  Acta Phys.\ Polon.\ B {\bf 40}, 2507 (2009)
  
%\cite{Barish:1977qk}
\bibitem{ref:anlccqe} 
  S.~J.~Barish {\it et al.},
  %``Study of Neutrino Interactions in Hydrogen and Deuterium. 1. Description of the Experiment and Study of the Reaction Neutrino d --> mu- p p(s),''
  Phys.\ Rev.\ D {\bf 16}, 3103 (1977).
  %doi:10.1103/PhysRevD.16.3103
  %%CITATION = doi:10.1103/PhysRevD.16.3103;%%
  %211 citations counted in INSPIRE as of 02 Nov 2016

\bibitem{ref:anlccres1} 
 Radecky et al. Phys Rev D, 3rd series, volume 25, number 5, 1 March 1982, p 1161-1173
 
 \bibitem{ref:anlccres2}
Thpr: Derrick et al. Phys Rev D, Vol 23, Number 3, 1 Feb 1981, p 569-575

%\cite{Baker:1981su}
\bibitem{ref:bnlccqe} 
  N.~J.~Baker {\it et al.},
  %``Quasielastic Neutrino Scattering: A Measurement of the Weak Nucleon Axial Vector Form-Factor,''
  Phys.\ Rev.\ D {\bf 23}, 2499 (1981).
  %doi:10.1103/PhysRevD.23.2499
  %%CITATION = doi:10.1103/PhysRevD.23.2499;%%
  %233 citations counted in INSPIRE as of 02 Nov 2016

\bibitem{ref:bnlccres1} 
Kitagaki et al. Phys Rev D, Vol 34, Number 9, 1 November 1986, p 2554-2565

\bibitem{ref:bnlccres2}
 K. Furuno et al., %Proceedings of the Second International Workshop on Neutrino-Nucleus Interactions in the Few-GeV Region 
 (NuInt02), UC Irvine, U.S.A., KEK Preprint 2003-48 
 
%\cite{Kitagaki:1983px}
\bibitem{ref:fnalccqe} 
  T.~Kitagaki {\it et al.},
  %``High-Energy Quasielastic Muon-neutrino n ---> mu- p Scattering in Deuterium,''
  Phys.\ Rev.\ D {\bf 28}, 436 (1983).
  %doi:10.1103/PhysRevD.28.436
  %%CITATION = doi:10.1103/PhysRevD.28.436;%%
  %188 citations counted in INSPIRE as of 02 Nov 2016
  
 %\cite{Allasia:1990uy}
\bibitem{ref:bebcccqe} 
  D.~Allasia {\it et al.},
  %``Investigation of exclusive channels in neutrino / anti-neutrino deuteron charged current interactions,''
  Nucl.\ Phys.\ B {\bf 343}, 285 (1990).
  %doi:10.1016/0550-3213(90)90472-P
  %%CITATION = doi:10.1016/0550-3213(90)90472-P;%%
  %71 citations counted in INSPIRE as of 02 Nov 2016
  
  \bibitem{ref:llewellynsmith}
Smith, CH Llewellyn
%``Neutrino reactions at accelerator energies,''
Physics Reports 3 5 1972

\bibitem{reinseghal} D.~Rein and L.~M.~Sehgal,
  %``Neutrino Excitation of Baryon Resonances and Single Pion Production,''
  Annals Phys.\  {\bf 133}, 79 (1981).

\bibitem{nubros}
  L.~Pickering, P.~Stowell and J.~Sobczyk,
  %``Event reweighting with the NuWro neutrino interaction generator,''
  arXiv:1610.07053 [hep-ex].
  %%CITATION = ARXIV:1610.07053;%%

\bibitem{ref:singh}
  S.~K.~Singh,
  %``The Effect of final state interactions and deuteron binding in neutrino d --> Mu- p p,''
  Nucl.\ Phys.\ B {\bf 36}, 419 (1972).
  %doi:10.1016/0550-3213(72)90227-1
  %%CITATION = doi:10.1016/0550-3213(72)90227-1;%%
  %36 citations counted in INSPIRE as of 02 Nov 2016


\bibitem{smithmoniz}R.~A.~Smith and E.~J.~Moniz,
  %``Neutrino Reactions On Nuclear Targets,''
  Nucl.\ Phys.\ B {\bf 43}, 605 (1972)
  
 \bibitem{nieves} J.~Nieves, I.~Ruiz Simo and M.~J.~Vicente Vacas,
  %``Inclusive Charged--Current Neutrino--Nucleus Reactions,''
  Phys.\ Rev.\ C {\bf 83}, 045501 (2011)
 
 \bibitem{temmodel} 
  A.~Bodek, H.~S.~Budd and M.~E.~Christy,
  %``Neutrino Quasielastic Scattering on Nuclear Targets: Parametrizing Transverse Enhancement (Meson Exchange Currents),''
  Eur.\ Phys.\ J.\ C {\bf 71}, 1726 (2011)
  %doi:10.1140/epjc/s10052-011-1726-y
  %[arXiv:1106.0340 [hep-ph]].
  %%CITATION = doi:10.1140/epjc/s10052-011-1726-y;%%
  %100 citations counted in INSPIRE as of 02 Nov 2016
  
  
\bibitem{ref:mbnu}
  A.~A.~Aguilar-Arevalo {\it et al.} [MiniBooNE],
  %``First Measurement of the Muon Neutrino Charged Current Quasielastic Double Differential Cross Section,''
  Phys.\ Rev.\ D {\bf 81}, 092005 (2010)
 % doi:10.1103/PhysRevD.81.092005
  %[arXiv:1002.2680 [hep-ex]].
  %%CITATION = doi:10.1103/PhysRevD.81.092005;%%
  %357 citations counted in INSPIRE as of 02 Nov 2016

\bibitem{ref:minnu} 
  G.~A.~Fiorentini {\it et al.} [MINERvA],
  %``Measurement of Muon Neutrino Quasielastic Scattering on a Hydrocarbon Target at Eν∼3.5  GeV,''
  Phys.\ Rev.\ Lett.\  {\bf 111}, 022502 (2013)
 % doi:10.1103/PhysRevLett.111.022502
 % [arXiv:1305.2243 [hep-ex]].
  %%CITATION = doi:10.1103/PhysRevLett.111.022502;%%
  %154 citations counted in INSPIRE as of 02 Nov 2016
  
\bibitem{ref:mbnub} 
  A.~A.~Aguilar-Arevalo {\it et al.} [MiniBooNE],
  %``First measurement of the muon antineutrino double-differential charged-current quasielastic cross section,''
  Phys.\ Rev.\ D {\bf 88}, no. 3, 032001 (2013)
 % doi:10.1103/PhysRevD.88.032001
 % [arXiv:1301.7067 [hep-ex]].
  %%CITATION = doi:10.1103/PhysRevD.88.032001;%%
  %113 citations counted in INSPIRE as of 02 Nov 2016

\bibitem{ref:minnub} 
  L.~Fields {\it et al.} [MINERvA],
  %``Measurement of Muon Antineutrino Quasielastic Scattering on a Hydrocarbon Target at Eν∼3.5  GeV,''
  Phys.\ Rev.\ Lett.\  {\bf 111}, no. 2, 022501 (2013)
  %doi:10.1103/PhysRevLett.111.022501
  %[arXiv:1305.2234 [hep-ex]].
  %%CITATION = doi:10.1103/PhysRevLett.111.022501;%%
  %129 citations counted in INSPIRE as of 02 Nov 2016
  
\bibitem{ref:hesse} 
J., F., and M. I. N. U. I. T. Roos. 
%"Minuit-a system for function minimization and analysis of the parameter errors and correlations." 
Comp. Physics Comm. 10.6 (1975): 343-367.



\bibitem{minervadata} P.~A.~Rodrigues {\it et al.} [MINERvA Collaboration],
  %``Identification of nuclear effects in neutrino-carbon interactions at low three-momentum transfer,''
  Phys.\ Rev.\ Lett.\  {\bf 116}, 071802 (2016)
  
%\bibitem{geniegen}  C.~Andreopoulos {\it et al.},
 % %``The GENIE Neutrino Monte Carlo Generator,''
  %Nucl.\ Instrum.\ Meth.\ A {\bf 614}, 87 (2010)





  





\end{thebibliography}
\end{document}